\documentclass{aa}
\usepackage{graphicx}
\def\lesssim{\mathrel{\mathpalette\fun <}}
\def\gtrsim{\mathrel{\mathpalette\fun >}}
\def\fun#1#2{\lower3.6pt\vbox{\baselineskip0pt\lineskip.9pt
\ialign{$\mathsurround=0pt#1\hfil##\hfil$\crcr#2\crcr\sim\crcr}}}

\begin{document}
   \title{Nucleosynthesis in $\gamma-$ray bursts outflows}


   \author{Martin Lemoine
          }

   \offprints{M. Lemoine}

   \institute{Institut d'Astrophysique de Paris, GReCO, C.N.R.S.,\\ 
98 bis boulevard Arago, F-75014 Paris, France\\ \email{lemoine@iap.fr} }

   \date{\today}

   \abstract{ It is shown that fusion of neutrons and protons to
   $^4$He nuclei occurs in $\gamma-$ray burst outflows in a process
   similar to big-bang nucleosynthesis in the early Universe. Only the
   surviving free neutrons can then decouple kinematically from the
   charged fluid so that the multi-GeV neutrino signal predicted from
   inelastic nuclear n$-$p collisions is significantly reduced. It is
   also argued that a sizeable fraction of ultra-high energy cosmic
   rays accelerated in $\gamma-$ray bursts should be $^4$He nuclei.
    }

   \maketitle

\section{Introduction. } 

Multi-wavelength observations of $\gamma-$ray bursts in the past
decade have given increasing evidence in favor of the ``expanding
fireball'' model (Paczy\'nski 1986, 1990; Goodman 1986, Shemi \& Piran
1990; Piran 2000 and M\'esz\'aros 2002 for reviews), in which a photon
-- pair plasma loaded with a small admixture of baryons expands
relativistically and converts the initial energy into baryon kinetic
energy. In the internal/external shock scenario (Rees \& M\'esz\'aros
1992; Rees \& M\'esz\'aros 1994; Piran 2000 and M\'esz\'aros 2002)
this kinetic energy is dissipated in shocks, giving rise to the
$\gamma-$ray burst phenomenon.  The identity of the inner engine, the
source of energy and the mechanism of collimation remain however
unknown.

One should not expect that only protons are injected in the
accelerating wind. As a matter of fact, all theoretical proposals for
the progenitor involve compact objects, {\it e.g.} neutron stars/black
hole mergers, or imploding massive stars, in which the baryon load
must be neutron rich. This has triggered recent interest in the study
of phenomenological consequences of neutron loading in $\gamma-$ray
bursts (e.g. Derishev {\it et al.}  1999a,b; Bahcall \& M\'esz\'aros
2000; Fuller {\it et al.}  2000; M\'esz\'aros \& Rees 2000).

One possible consequence is nucleosynthesis of neutrons and protons to
heavier nuclei.  A recent study argued that nucleosynthesis should not
take place as the dynamical timescale (taken to be $t_{\rm
dyn}\approx10^{-5}\,$sec) is too short (Derishev {\it et al.}
1999). However this particular value rather constitutes a strict lower
limit to the dynamical timescale since it implies a source size
$\lesssim 3\,$km. This timescale is nonetheless bounded from above by
the shortest variability timescale observed $\approx10^{-2}\,$sec
(Piran 1999).

The purpose of this {\it Letter} is to show that nucleosynthesis to
$^4$He can occur and is generically efficient, provided the dynamical
timescale $\gtrsim 10^{-4}\,$sec (Section~\ref{nucleosynthesis}). Most
available neutrons and protons are then bound into $^4$He and only the
surviving neutrons can decouple kinematically from the charged fluid
component (Section~\ref{discussion}). Thus this significantly weakens
the $5-15\,$GeV neutrino signal expected from $n-p$ inelastic
collisions (Derishev {\it et al.} 1999a; Bahcall \& M\'esz\'aros 2000;
M\'esz\'aros \& Rees 2000). Furthermore, if $\gamma-$ray bursts
accelerate particles to ultra-high energies $\gtrsim10^{20}\,$eV
(Levinson \& Eichler 1993; Vietri 1995; Waxman 1995; Waxman 2001 for a
review), the ultra-high energy cosmic ray spectrum should comprise a
sizeable fraction of $^4$He nuclei (Section~\ref{discussion}).

\section{Nucleosynthesis.} \label{nucleosynthesis}

The fireball wind can be modeled as a pair plasma with luminosity $L=
L_{50}\times10^{50}\,$ergs/s injected into a solid angle
$\Omega=4\pi\theta^2$, where
$\theta^2\equiv\left(1-\cos\theta_{1/2}\right)$ represents the effect
of beaming into two jets of half-opening angle
$\theta_{1/2}$. Observations suggest a roughly uniform $L_{50}\sim1$
among $\gamma-$ray bursts and a varying $\theta$ with mean value
$\theta\sim0.1$ (Frail {\it et al.} 2001; Panaitescu \& Kumar 2001;
Piran {\it et al.} 2001); for a wind of typical duration
$\sim10\,$sec, this yields a total equivalent isotropic output energy
$\sim 10^{53}\,$ergs. The bulk Lorentz factor is written $\Gamma$, and
its saturation value $\eta\equiv L/\dot{M}c^2$ is the ratio of total
luminosity to baryon outflow (M\'esz\'aros {\it et al.}
1992). Injection takes place at radius $r_o=r_{o,7}\times10^7\,$cm,
with initial temperature $T_0\simeq
0.93\,L_{50}^{1/4}\theta^{-1/2}_{0.1}r_{o,7}^{-1/2}\,$MeV$/k_{\rm B}$,
using $11/2$ for the number of degrees of freedom (photons +
pairs). In the wind frame the ejecta is indeed similar to the early
Universe before big-bang nucleosynthesis (Shemi \& Piran
1990). However the physical conditions are very different: in a
$\gamma-$ray burst ejecta, the comoving timescale $\sim 10^{-3}\,$sec
and the photon to baryon ratio $n_\gamma/n_{\rm b}\simeq 4.1\times
10^4 L_{50}^{-1/4}\theta^{1/2}_{0.1}r_{o,7}^{1/2}\eta_{300}$, with
$\eta_{300}\equiv\eta/300$, to be compared with $t_{\rm
dyn}\sim100\,$sec and $n_\gamma/n_{\rm b}\sim10^{10}$ for the early
Universe. A large dynamical timescale favors nucleosynthesis but a
high entropy acts against. Big-bang nucleosynthesis is also hampered
by a small neutron to proton ratio due to neutron decay and late
freeze-out of the weak interactions that interconvert protons and
neutrons. Here one expects equal $n, p$ mass fractions $X_{\rm n}\sim
X_{\rm p}$ if baryons come from photodissociated nuclei, and neutron
decay is insignificant on a millisecond timescale. If $T_0\gtrsim
6.5{\rm MeV}\, (t_{\rm dyn}/10^{-3}{\rm sec})^{-1/5}$, the rate of
weak interactions becomes larger than the fireball expansion rate and
the neutron to proton ratio achieves equilibrium independently of the
initial composition, $X_{\rm n}/X_{\rm p}\rightarrow\exp[(m_{\rm
p}-m_{\rm n})c^2/k_{\rm B}T_0]\sim1$. However this occurs for
$\theta\lesssim 0.02r_{o,6}^{-1}L_{50}^{1/2}(t_{\rm dyn}/10^{-3}{\rm
sec})^{2/5}$, {\it i.e.} for the most highly beamed or compact
$\gamma-$ray bursts, or those with the longest dynamical timescales
(see also Fuller {\it et al.} 2000 for a discussion of the impact of a
neutrino flux on the initial chemical composition of the outflow).

Adiabaticity during the acceleration phase implies that the bulk
Lorentz factor $\Gamma$ and the temperature $T$ evolve as $T\propto
\Gamma^{-1}\propto r^{-1}$, with $r$ the radial coordinate in the lab
frame. If the flow can be approximated as quasi-one dimensional, the
relativistic Bernoulli equation further implies $\Gamma\propto\theta
r$ (Blandford \& Rees 1974). In particular M\'esz\'aros {\it et al.}
(1993) use $\Gamma\approx\theta r/r_o$, which is obviously valid only
in the limit $r\gg \theta^{-1}r_o$. The exact numerical prefactor in
this relation depends on the details of the injection phase at radius
$r\sim r_o$.  In what follows, the early behavior will be
phenomenologically modeled as $\Gamma=(1 + \theta r/r_o)$, with
$T\propto \Gamma^{-1}$, which approaches the twin exhaust model of
Blandford \& Rees (1974), in which the flow is collimated by outside
pressure forces and accelerates to relativistic velocity through a de
Laval nozzle. A more exact solution requires solving the complex
problem of injection and collimation in the early phase of the
flow. Steps in this direction have been accomplished recently by
Levinson \& Eichler (2000) who studied the hydrodynamic collimation of
a $\gamma-$ray burst outflow by a wind emanating from a torus. Their
solution indeed reproduces a nozzle with nearly constant cross-section
at small radii similarly to the Blandford \& Rees (1974) model.

 The dependence of the bulk Lorentz factor on $r$ is important as it
gives the dynamical expansion timescale $t_{\rm dyn}\equiv r/\Gamma
c$. Interestingly in this context the scaling $\Gamma \propto \theta
r$ suggests that highly beamed outflows have a longer dynamical
timescale, hence should not show variations on short timescales due to
erasure of inhomogeneities on scales smaller than the sound horizon
$\lesssim c t_{\rm dyn}/\sqrt{3}$.  Nevertheless, in order to
circumvent uncertainties related to the modeling of the early
evolution of the bulk Lorentz factor, all results that follow will be
shown as a function of dynamical timescale $t_{\rm dyn}$ instead of
$\Gamma$. This is justified as $t_{\rm dyn}$ and the entropy are the
two parameters that control the efficiency of
nucleosynthesis. Moreover it was checked that a spherically symmetric
wind with the simple expansion law $\Gamma=r/r_o$ gives the same
result than a jet with respect to nucleosynthesis provided the
dynamical timescale and entropy are the same.

In the ultra-relativistic regime $\Gamma\gg1$ the temperature
decreases exponentially fast with comoving time
$\propto\log(r/r_o)$. In the outflow thermal equilibrium is ensured
between all species all along nucleosynthesis, and neutrons remain
coupled to protons through nuclear scattering with velocity averaged
cross-section $\langle\sigma_{\rm n-p} v\rangle\sim30$mb~c (Derishev
{\it et al.} 1999b). Decoupling occurs when $1/n_{\rm
p}\langle\sigma_{\rm n-p}v\rangle > t_{\rm dyn}$, at temperature
$T\approx 2\,
L_{50}^{-1/12}\theta_{0.1}^{1/6}r_{o,7}^{1/6}\eta_{300}^{1/3}(t_{\rm
dyn}/10^{-3}{\rm sec})^{-1/3}\,{\rm keV}/k_{\rm B}$, {\it i.e.}, well
after nucleosynthesis has taken place (see below).

\begin{figure}
   \centering \includegraphics[width=0.5\textwidth]{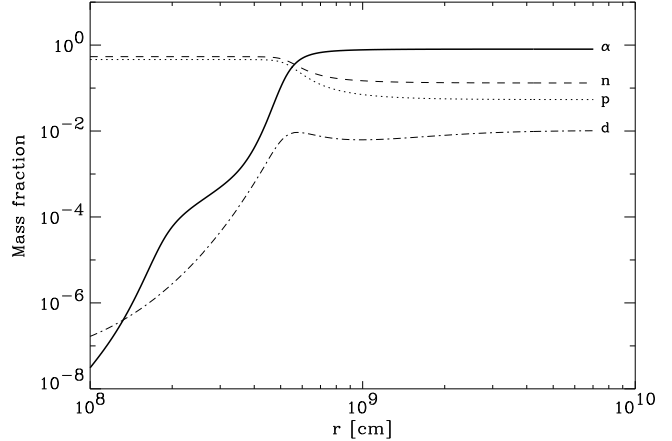}
\caption{Mass fractions {\it vs.} radius for the case
$r_{o}=3\times10^6\,$cm, $\theta=0.1$ ($\Omega/4\pi=10^{-2}$),
dynamical timescale $t_{\rm dyn}=10^{-3}\,$sec,
$L=10^{50}\,$ergs.s$^{-1}$ and initial mass composition
$X_{\rm n}=0.54$, $X_{\rm p}=0.46$.} \label{f1}
\end{figure}

In order to estimate the final abundances of nuclei synthesized, use
was made of a big-bang nucleosynthesis numerical code whose time and
entropy evolutions were modified accordingly. This code accounts
successfully for synthesis of elements up to $^7$Be when compared to
other big-bang nucleosynthesis calculations. An example of the outcome
of nucleosynthesis is shown in Fig.~\ref{f1} for a $\gamma-$ray burst
with fiducial parameters $r_{o,7}=0.3$, $\theta=0.1$, $L_{50}=1$,
initial composition $X_{\rm n}=0.54$, $X_{\rm p}=0.46$ (photodissociated
$^{56}$Fe nuclei) and dynamical timescale $t_{\rm
dyn}\simeq10^{-3}\,$sec.  The final mass fractions of elements
produced are $X_4\simeq0.84$ (for $^4$He), $X_{\rm n}\simeq0.12$,
$X_{\rm p}\simeq0.04$, $X_{\rm D}\simeq0.005$ and other elements are
produced only in traces. 
Although the numerical code used cannot deal with elements beyond mass
9, estimates of the triple $\alpha$ reaction and $\alpha(\alpha\,
n,\gamma)^9$Be$(\alpha,n)^{12}$C necessary to bridge the mass gaps to
$^{12}$C and beyond suggest that these three-body reactions do not
have sufficient time to produce heavy elements in abundance (see
however Section 3).

\begin{figure}
   \centering \includegraphics[width=0.5\textwidth]{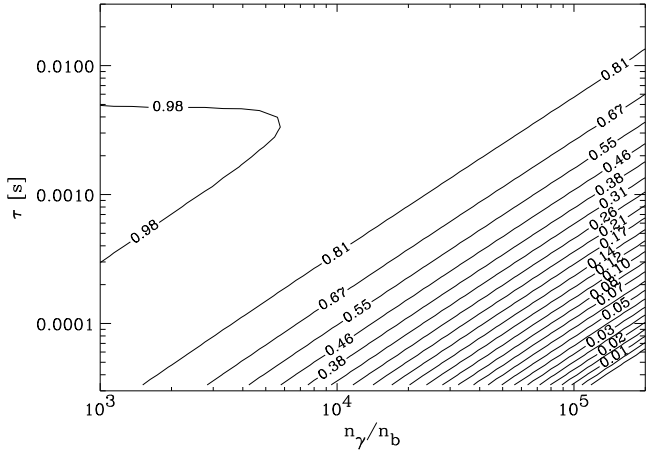}
\caption{Contours of the final $^4$He mass fraction as a function of
photon to baryon number density $n_\gamma/n_{\rm b}$ and comoving
dynamical timescale $\tau_{\rm dyn}$.} \label{f2}
\end{figure}

The bulk of nucleosynthesis occurs at temperature $T_{\rm
nuc}\sim0.1-0.2\,$MeV$/k_{\rm B}$, corresponding to $\Gamma\sim
10-100$ depending on the parameters, and is over by $T_{\rm
nuc}\sim0.06\,$MeV$/k_{\rm B}$. Nucleosynthesis is favored by a large
$t_{\rm dyn}$ and a small $n_{\gamma}/n_{\rm b}$, and it is thus more
efficient as $\eta$ decreases (meaning higher baryon density), as
$\theta$ decreases, or as $r_{o}$ increases if $t_{\rm dyn}\propto
r_o$ (since a large expansion timescale is more critical to
nucleosynthesis than a small entropy). Obviously the efficiency of
nucleosynthesis is maximal for $X_{\rm n}/X_{\rm p}=1$ initially, and less
efficient as this ratio departs from unity. A contour plot of the
final mass fraction of $^4$He produced is shown in Fig.~\ref{f2} as a
function of $t_{\rm dyn}$ and $n_\gamma/n_{\rm b}$. In this figure, it
was assumed that initially $X_{\rm n}/X_{\rm p}=1$.  It shows that
nucleosynthesis is efficient in most of parameter space: most
available neutrons and protons are bound into $^4$He provided the
dynamical timescale $t_{\rm dyn}\gtrsim10^{-4}\,$sec if
$n_\gamma/n_{\rm b}\lesssim 10^4$, or $t_{\rm dyn}\gtrsim10^{-3}\,$sec
if $n_\gamma/n_{\rm b}\lesssim 10^5$.  One can understand analytically
these results by comparing the expansion rate $t_{\rm dyn}^{-1}$ to
the deuterium formation rate at temperature $T$ ($n_{\rm b}\propto
T^{-3}$): $R_{{\rm p}({\rm n},\gamma){\rm d}}=\langle\sigma_{{\rm
p}({\rm n},\gamma){\rm d}}v\rangle X_{\rm n}(1-X_{\rm n})n_{\rm b}
\simeq 3\times 10^4 T_{0.1{\rm
MeV}}^{-3}r_{o,7}^{-1/2}\theta_{0.1}^{-1/2}L_{50}^{1/4}\eta_{300}^{-1}
X_{\rm n}(1-X_{\rm n})\,{\rm sec}^{-1}$. Indeed deuterium production
occurs if $t_{\rm dyn}\gtrsim 10^{-4}\,$sec. The dependence on the
entropy comes through the reverse photodisintegration rate which
inhibits the production of deuterium hence of heavier elements.

One expects nucleosynthesis to be more efficient than above if, for
instance, magnetic energy represents a significant fraction $x_B$ of
the total energy. The photon number density hence $n_\gamma/n_{\rm b}$
are then smaller by $(1-x_B)^{3/4}$ while the initial temperature is
only reduced by $(1-x_B)^{1/4}$.  The present calculation also assumes
a homogeneous steady wind, but inhomogeneities in the wind,
e.g. shells or blobs, would increase the efficiency of
nucleosynthesis, as $n_\gamma/n_{\rm b}\propto E^{-1/4}$ where $E$ is
the local energy content.

\section{Discussion.}\label{discussion}

The feedback of nuclear binding energy release on the fireball
evolution has been neglected in the above calculations. This is
justified since the ratio of entropy density released to total entropy
density (assuming this latter is dominated by the photons) if all
neutrons and protons are instantaneously converted to $^4$He at
$T\approx0.1\,$MeV$/k_{\rm B}$ reads ${\rm d}S/S \approx 20 (k_{\rm
B}T/0.1{\rm MeV})^{-1} (n_\gamma/n_{\rm b})^{-1}$, with ${\rm
d}S=n_{\rm b}B_4/4k_{\rm B}T$ and $B_4=28.3\,$MeV the $^4$He binding
energy.  Note that in some rather extreme parts of parameter space,
the prerequisite conditions for successful $r-$process nucleosynthesis
could be satisfied: initial electron fraction $Y_{\rm e}\equiv X_{\rm
p}<0.5$, $t_{\rm dyn}\gtrsim10^{-3}\,$sec, $n_\gamma/n_{\rm
b}\sim10^2$ and initial temperature $T_0\sim 1\,$MeV$/k_{\rm B}$
(e.g. Hoffman {\it et al.}  1997; Meyer \& Brown 1997). This
presumably requires a highly beamed ejecta $\theta < 0.1$ from a
compact source $r_o\lesssim 10^6\,$cm, with a significant fraction of
magnetic energy, and/or a low limiting Lorentz factor $\eta$. This
latter condition can be fulfilled in the outer parts of a jet with
inhomogeneous baryon load as proposed recently (Zhang \& M\'esz\'aros
2002). Successful $r-$process nucleosynthesis might thus be able to
operate in $\gamma-$ray bursts outflows, and further studies using
dynamical $r-$process codes appear mandatory. Other connections
between the site $r-$process site and $\gamma-$ray bursts have been
proposed by Eichler {\it et al.} (1989), Levinson \& Eichler (1993)
and Cameron (2001).

An important consequence of successful nucleosynthesis to $^4$He is to
keep neutrons tied to protons and prevent their kinematical decoupling
when nuclear scattering becomes ineffective. As shown by Bahcall \&
M\'esz\'aros (2000) and M\'esz\'aros \& Rees (2000), such decoupling
occurs before the Lorentz factor has saturated to its limiting value
$\eta$ provided $\eta\gtrsim 480
L_{50}^{1/4}\theta_{0.1}^{-1/2}r_{o,7}^{-1/4}Y_{\rm e}^{1/4}$ and leads to
relative velocities between neutrons and protons $\sim c$; in turn
this leads to pion production in inelastic collisions and thus to a
$5-15\,$GeV neutrino signal.  M\'esz\'aros \& Rees (2000) have further
shown that transverse diffusion of neutrons in inhomogeneously baryon
loaded fireballs can lead to an appreciable multi-GeV neutrino signal
for lower values of $\eta$. However these studies assumed that
nucleosynthesis did not occur. In fact, the inelastic collisions occur
in these scenarii at large radii well after nucleosynthesis, hence
only the surviving free neutrons will be able to decouple.

Consider first an homogeneous fireball. In the absence of
nucleosynthesis, the neutrino signal is proportional to
$X_{\rm p}X_{\rm n}\equiv Y_{\rm e}\left(1-Y_{\rm e}\right)$, and is thus maximal when
$Y_{\rm e}\sim0.5$. The neutron$-^4$He collision cross-section is higher
than the $n-p$ cross-section by the geometrical factor $4^{2/3}$ but
there are only $X_4/4$ $^4$He nuclei per baryon ($X_4$ denotes as
before the mass fraction). The neutrino signal produced after
nucleosynthesis ($X_4\neq0$) is then a factor $r$ of that produced
when $X_4=0$ (no nucleosynthesis) with:

$$r\simeq \frac{2\left(1-Y_{\rm e}\right)-X_4}{4Y_{\rm
e}\left(1-Y_{\rm e}\right)}\left(2Y_{\rm e} - 0.685X_4\right);$$

  it was assumed that the outcome of nucleosynthesis is only $n$, $p$
and $^4$He. This reduction factor takes the following values: for
$Y_{\rm e}\simeq0.5$, $r\simeq0.33$ for $X_4=0.5$, $r\simeq0.09$ for
$X_4=0.80$. If $Y_{\rm e}>0.5$ and nucleosynthesis is maximally efficient,
all neutrons are bound into $^4$He and obviously $r=0$. By combining
the above relation with Fig.~\ref{f2} it is possible to obtain
estimates of $r$ for various $\gamma-$ray bursts parameters. As an
example, $\eta=500$ (so that kinematical decoupling of free neutrons
occurs before saturation), $L=10^{50}\,$ergs$\,$s$^{-1}$,
$\theta=0.1$, $r_o=0.3\times10^7\,$cm and $t_{\rm dyn}=10^{-3}\,$sec
gives $X_4\simeq0.70$ and the neutrino signal is reduced by a factor
$\approx6.4$. Note that even a small reduction factor is significant
as the neutrino signal predicted is of the order a few events per year
for a km$^3$ neutrino telescope and does not exceed a dozen events per
year.

  If the jet is inhomogenous, say made of an inner baryon poor jet
surrounded by a baryon rich outer shell (Eichler \& Levinson 1999,
M\'esz\'aros \& Rees 2001), the above discussion still applies if the
outer baryonic wind originates from the central engine with a
temperature $T\sim\,{\cal O}({\rm MeV})$ similar to that of the
central jet. In effect nucleosynthesis is then very efficient in the
outer shell since its Lorentz factor and entropy are both lower than
in the jet. If the surrounding shell is ``cold'', or if it is neutron
rich and the inner jet proton rich (or vice-versa) one can circumvent
the above argument. However in the case of a jet punching its way
through a collapsar progenitor atmosphere M\'esz\'aros \& Rees (2000)
have shown that it leads to a very low neutrino signal.

Finally, if ultra-high energy cosmic rays are accelerated in
$\gamma-$ray bursts (Levinson \& Eichler 1993; Vietri 1995; Waxman
1995; Waxman 2001 and references), one expects in the present context
a significant fraction of these particles to be $^4$He nuclei. However
nuclei are subject to photodisintegration with cross-section
$\sim$a~few~mb for photon energies $\gtrsim20\,$MeV in the nucleus
rest frame, and during acceleration in internal shocks and production
of the $\gamma-$ray signal a fraction of them will be disrupted.

The calculation of the fraction of nuclei dissociated as a function of
their energy and shock radius is in itself similar to that performed
by Waxman \& Bahcall (1997), Guetta {\it et al.} (2001) for the
production of a 100~TeV neutrino signal from pion production of
accelerated protons. At the highest energies $E_{{\rm
He}}\gtrsim10^{15}\,$eV (observer frame), one can thus write the
photodisintegration rate $R\equiv A^{-1}{\rm d}A/{\rm d}t\simeq
U_\gamma c \eta \langle\sigma_{\gamma{\rm He}}\rangle/ A\epsilon_{\rm
b}$, where $U_\gamma\equiv L_\gamma / 4\pi\theta^2 r^2\eta^2c$ is the
photon energy density, $\epsilon_{\rm b}\approx 1\,$MeV is the
observed break energy in the $\gamma-$ray spectrum, and
$\langle\sigma_{\gamma{\rm He}}\rangle$ is the inverse energy weighted
photodisintegration cross-section, $\langle\sigma_{\gamma{\rm
He}}\rangle\simeq 2.8\,$mb accounting for 1 and 2 nucleon loss with
respective branching ratios 80\%, 20\%.  The optical depth to
photodisintegration thus reads $\tau\approx Rt_{\rm dyn}\approx
L_{50}\theta_{0.1}^{-2}r_{14}^{-1}\eta_{300}^{-2}(\epsilon_{\rm
b}/1\,{\rm MeV})^{-1}$, with $r_{14}\equiv r/10^{14}\,{\rm cm}$ the
radius of emission.

Since features of temporal width $\delta t\equiv\delta t_{-2}\times
10^{-2}\,$sec are emitted at radius $r_{14}\simeq0.6\delta
t_{-2}\eta_{300}^2$, this optical depth is unity where features of
width $\delta t\sim10^{-2}\,$sec are emitted. High energy nuclei
accelerated in shocks at smaller radii are photodisintegrated, while
those accelerated at larger radii are unharmed. One can also show that
low energy nuclei are not photodisintegrated even at small radii. The
overall $^4$He nuclei energy spectrum is thus subject to
photodisintegration, acceleration and reacceleration in shocks and
adiabatic losses.  Acceleration at large radii is in any case crucial
to overcome expansion losses (Waxman 2001). It thus seems reasonable
to expect that a sizeable fraction of $^4$He nuclei would be present
in the escaping ultra-high energy radiation and the measurement of the
chemical composition might provide further information on the nuclear
processes at work in the $\gamma-$ray burst. Detailed signatures of
these ultra-high energy $^4$He nuclei will be presented elsewhere.
\bigskip

\noindent{\it Note added:} While this paper was being refereed, a
similar study by Pruet {\it et al.} (2002) appeared, with similar
conclusions as to the efficiency of nucleosynthesis.

\begin{acknowledgements}
It is a pleasure to thank R.~Mochkovitch, G.~Pelletier and E.~Waxman
for useful comments and suggestions.
\end{acknowledgements}

\end{document}